\documentclass[conference]{IEEEtran}
\usepackage{cite}
\usepackage{amsmath,amssymb,amsfonts}
\usepackage{algorithmic}
\usepackage{graphicx}
\usepackage{textcomp}
\usepackage{xcolor}
\usepackage{tikz}

\renewcommand{\circ}{\,\raisebox{1.5pt}{\tikz \draw[line width=0.4pt] circle(0.9pt);}\,}

\def\BibTeX{{\rm B\kern-.05em{\sc i\kern-.025em b}\kern-.08em
    T\kern-.1667em\lower.7ex\hbox{E}\kern-.125emX}}

\renewcommand{\IEEEQED}{\IEEEQEDopen}

\newtheorem{definition}{Definition}[section]
\newtheorem{theorem}[definition]{Theorem}
\newtheorem{remark}[definition]{Remark}
\newtheorem{corollary}[definition]{Corollary}

\newcounter{abccnt}
\newenvironment{abc}{\begin{list}{\bf(\alph{abccnt})}{\usecounter{abccnt}
\labelwidth4ex \labelsep1ex \leftmargin6ex
\parsep3pt \itemsep1pt \topsep3pt}}{\end{list}}

\begin{document}

\title{Group testing via residuation and partial geometries}

\author{\IEEEauthorblockN{Marcus Greferath and Cornelia R\"o\ss{}ing}
\IEEEauthorblockA{\textit{School of Mathematics and Statistics} \\
\textit{University College Dublin}\\
Dublin, Republic of Ireland \\
marcus.greferath@ucd.ie, roessing@maths.ucd.ie}
}

\maketitle

\begin{abstract} The motivation for this paper comes from the ongoing SARS-CoV-2 Pandemic. Its goal is to present a previously neglected approach to non-adaptive group testing and describes it in terms of residuated pairs on partially ordered sets. Our investigation has the advantage, as it naturally yields an efficient decision scheme (decoder) for any given testing scheme. This decoder allows to detect a large amount of infection patterns.

Apart from this, we devise a construction of good group testing schemes that are based on incidence matrices of finite partial linear spaces. The key idea is to exploit the structure of these matrices and make them available as test matrices for group testing. These matrices may generally be tailored for different estimated disease prevalence levels. As an example, we discuss the group testing schemes based on generalized quadrangles. 

In the context at hand, we state our results only for the error-free case so far. An extension to a noisy scenario is desirable and will be treated in a subsequent account on the topic.
\end{abstract}

\smallskip


\section{Introduction}
During the initial low-prevalence phase of a pandemic like the current SARS-CoV-2 pandemic, where a new pathogen has started spreading, it is helpful to use a technique called {\em group testing\/} in order to identify infected individuals, particularly when testing is complicated or otherwise expensive. 

The technique goes back to initial ideas by Dorfman~\cite{dorfman43} in 1943, whose challenge was to test WWII troops for syphilis based on an insufficient number of tests available. As the rate of infected soldiers was rather small (far below $10\%$), it turned out to be possible to significantly reduce the number of tests exploiting the following idea: Test a group of samples in a way that each test is used for a pool of specimen from a variety of samples. Dorfman correctly observed, that it was possible to identify a small number of infected participants out of a larger batch without having to medically check each individual for the infection in question. His method was further refined by Katona~\cite{katona73} who spread the specimen over several test. This method may be considered as a first approach to non-adaptive group testing.s 

To organize group tests we follow Katona~\cite{katona73} and divise a table (a binary matrix) for tests and participants showing which participant's specimen is contained in which test. For the test, an (unknown) vector of samples is multiplied by the test matrix using the arithmetic of the Boolean semi-field ${\mathbb B}_2$. The resuling test vector (syndrome) contains information about the infection pattern of the initial vector, and the goal is to reconstruct this pattern from the syndrome. 

In this paper, we show how this approach to group testing is closely related to the mathematical discipline of {\em Residuation Theory\/}. Its advantage is, that it comes with a natural choice of {\em decision scheme\/} (syndrome decoder) that generically works for a large class of infection patterns.

Adopting and refining conditions and techniques discussed in~\cite{essential}, we suggest to optimize the quality of the testing task by using incidence matrices of a class of finite geometries. These bear advantages due to certain intrinsic regularity, sparsity, and symmetry properties. 

We will state our results merely for the case where the tests are error-free, where an adjustment to noisy scenarios is devoted to a subsequent article. Our approach is clearly not tied to a particular pathogen, nor does the pathogen itself have to be viral. Its novelty lies in the (re-)establishment of an order-theoretic approach in conjunction with the application of finite geometries to design  non-adaptive pool testing strategies.

\section{Preliminaries}

\subsection{Linear Algebra on the Boolean Semifield}

In contrast with a set-based combinatorial approach to be found in~\cite{essential}, the Boolean semifield ${\mathbb B}_2$ will play a main role in this paper. On the $2$-element set $\{0,1\}$, it is the only non-trivial (yet meaningful) alternative to the binary field ${\mathbb F}_2$, its operations are described in Figure~\ref{d2op}. 

\begin{center}
\begin{figure}[h]
$$\begin{array}{|c||cc|}
\hline
+ & 0 & 1 \\
\hline\hline
0 & 0 & 1 \\
1 & 1 & 1\\
\hline
\end{array} \hspace{1cm} \begin{array}{|c||cc|}
\hline
\cdot & 0 & 1 \\
\hline\hline
0 & 0 & 0 \\
1 & 0 & 1\\
\hline
\end{array}$$
\caption{Operation tables of the boolean semifield ${\mathbb B}_2$}\label{d2op}
\end{figure}
\end{center}

\vspace{-5mm}

An interesting aspect of ${\mathbb B}_2$, that appears trivial on a first glance: it comes with an order relation $\leq$, naturally defined by $0 \leq 1$, and with a negation mapping ${\mathbb B}_2 \longrightarrow {\mathbb B}_2,\; x \mapsto \overline{x}$ where $\overline{0} =1$ and $\overline{1} =0$. This involution has the properties $\overline{x+y} = \overline{x}\,\overline{y}$ and $\overline{xy} = \overline{x} + \overline{y}$ which are well-known as {\em de~Morgan's\/} laws.

The commutative idempotent monoid ${\mathbb B}_2^n$ (equipped with componentwise addition $+$ and identity $\mathbf 0$) may be considered as a ${\mathbb B}_2$-vectorspace with $1\,x = x$ and $0\, x = \mathbf 0$ for all $x\in {\mathbb B}_2^n$. We have the distributive laws 
\begin{eqnarray*}
\lambda \, (x+y) & = & \lambda\, x + \lambda \, y\\
(\lambda + \mu)\, x & = & \lambda \, x + \mu \, x
\end{eqnarray*}
satisfied for all $\lambda, \mu \in {\mathbb B}_2$ and $x,y\in {\mathbb B}_2^n$. 

We will naturally extend order $\leq$ and negation to ${\mathbb B}_2^n$ and omit references to $n$ where-ever confusion is impossible.

The reader may readily recognize that what we are talking about is known from {\em Boolean Algebra\/}. The partial order on ${\mathbb B}_2^n$ is given via $x \leq y$ if and only if $y = x+y$.

In fact, defining multiplication $\cdot$ componentwise with identity $\mathbf 1$ on ${\mathbb B}_2^n$, we obtain the mentioned Boolean Algebra as the $6$-tuple $({\mathbb B}_2^n,+,\cdot,\mathbf 0, \mathbf 1, {\mathbb B}_2)$.

We say, the elements $b_1, \ldots, b_\ell \in {\mathbb B}_2^n$ be {\em linearly independent\/}, if $$b_i \not\leq \; \sum_{j\neq i} b_j, \quad \mbox{for all $1\leq i \leq \ell$}.$$

 The {\em Hamming distance\/} is defined as the function $$\delta: {\mathbb B}_2 \times {\mathbb B}_2 \longrightarrow {\mathbb N},\quad (x,y) \mapsto \left\{\begin{array}{lcl}
1 & : & x \neq y,\\
0 & : & \mbox{otherwise.}
\end{array}\right.$$ 

Following general conventions, we extend this function additively to ${\mathbb B}_2^n$ and obtain that $$\delta(x,y) \; = \; \left|\{i \in \{1, \ldots, n\} \mid x_i \neq y_i\}\right|.$$ 

In a similar fashion we define the Hamming weight $w: {\mathbb B}_2^n \longrightarrow {\mathbb N}$ such that $w(x) = |{\rm supp}(x)|$ for all $x\in {\mathbb B}_2^n$. As common in coding theory, we observe $w(x) = \delta(x,\mathbf 0)$, and, at least as important as the previous, $$\delta(x,y) = w(x+y) - w(x\cdot y).$$

The Hamming disk of radius $d$ centered in $x \in {\mathbb B}_2^n$ will be denoted by $$B_\delta(x,d)\; := \; \{z\in {\mathbb B}_2^n\mid \delta(x,z) \leq d\}.$$ Particularly the disks $B_\delta(\mathbf{0},d)$ and $B_\delta(\mathbf{1},d)$ will play a prominent role in our discussion.

\subsection{Two conditions}

The following concepts are slightly deviating from what is standard in the literature. For more information see~\cite{essential} and references given there.

\smallskip

\begin{definition}\label{sepanddis}
Let $H$ be an $n\times k$-matrix over the semifield ${\mathbb B}_2$. For a natural number $d$ consider the following two properties: 
\begin{description}\IEEEsetlabelwidth{d-Sep}\IEEEusemathlabelsep
\item[\bf d--Rev:] If $x\in B_\delta(\mathbf{0},d)$ and $y\in {\mathbb B}_2^n$, then $$xH\; = \; yH \;\; \mbox{will imply} \;\; x = y.$$
\item[\bf d--Dis:] For any $t\leq d$ and a set $T:=\{y_1, \ldots, y_t\}$ of rows of $H$ every row $x$ of $H$ not contained in $T$ satisfies $x\not\leq y_1+\ldots +y_t$.
\end{description}
\end{definition}

\smallskip

Property {\bf d--Rev} is a logically stronger version of {\em $d$-separability\/} introduced in~\cite{essential}, which says, that the mapping $H: {\mathbb B}_2^n \longrightarrow {\mathbb B}_2^k,\; x\mapsto xH$ is injective on $B_\delta(\mathbf{0},d)$. Property {\bf d--Dis} is called {\em $d$-disjunctness\/} in~\cite{essential}, and one of its immediate consequences is that any set of $d+1$ rows of $H$ is linearly independent in the sense defined above. We will show that conditions {\bf d--Rev} and {\bf d--Dis} are equivalent.

\smallskip

\begin{theorem}
Property {\bf d--Dis} implies property {\bf d--Rev}.
\end{theorem}
\begin{IEEEproof} Assume, the $n\times k$-matrix $H$ satisfies {\bf d--Dis}, and let $x\in B_\delta(\mathbf{0},d)$ and $y\in {\mathbb B}_2^n$ be given, such that $xH=yH$. If $x\neq y$, then w.l.o.g.~we may assume that there exists $i \in {\rm supp}(y)$ such that $i\not\in {\rm supp}(x)$. If not, then $y\in B_\delta(\mathbf{0},d)$ as well, and we only need to interchange the roles of $x$ and $y$. We conclude that row $h_i$ of $H$ satisfies $h_i \not\leq xH$ which comes from assumption {\bf d--Dis}. In contrast, $h_i \leq yH = xH$, a contradiction. This proves {\bf d--Rev} for $H$.
\end{IEEEproof}

\smallskip

\begin{theorem}
Property {\bf d--Rev} implies property {\bf d--Dis}.
\end{theorem}
\begin{IEEEproof} Assume, $H$ satisfies {\bf d--Rev}, and let $h_{i_1}, \ldots h_{i_t}$ be a sequence of $t$ rows of matrix $H$, where $t\leq d$. Then there is $x\in B_\delta(\mathbf{0},t) \subseteq B_\delta(\mathbf{0},d)$ such that $xH=\sum_{j=1}^t h_{i_j}$. For $\ell\not\in\{i_1, \ldots, i_t\}$ we set $y:=x + e_\ell$, where $e_\ell$ is the vector with only non-zero entry $1$ exactly in the $\ell$th position. If $h_\ell \leq \sum_{j=1}^t h_{i_j}$, then $xH=yH$ which forces $x=y$ by {\bf d--Rev}, a contradiction, because $x\neq y$. Hence $h_\ell \not\leq \sum_{j=1}^t h_{i_j}$, which is what condition {\bf d--Dis} asserts.  
\end{IEEEproof}

\subsection{Residuated Mappings}

In this section we present the main conceptual novelty of this paper. This will ease the discussion of group testing in that an efficient decision scheme is presented under quite general assumptions.

\smallskip

\begin{definition}\label{residuated}
Let $(A,\leq)$ and $(B,\leq)$ be two partially ordered sets. For mappings $f: A \longrightarrow B$ and $g: B \longrightarrow A$, the pair $(f,g)$ is called a {\em residuated pair\/}, if there holds $$f(x) \leq y \; \Longleftrightarrow \; x \leq g(y), \;\;\mbox{for all $x\in A$ and $y\in B$}.$$ 
\end{definition}

\smallskip

We will briefly collect basic facts about residuated pairs. For a source, the reader is referred to~\cite{blyth}. 

\smallskip

\begin{description}\IEEEsetlabelwidth{Fact:}\IEEEusemathlabelsep\itemsep=0.5mm
\item[Fact 1:] $f:A \longrightarrow B$ may be called a {\em residuated mapping\/}, if there is $g: B \longrightarrow A$, such that $(f,g)$ is a residuated pair. The mapping $g$ is then uniquely determined by $f$. Dually, $f$ is uniquely determined by $g$ which is called the {\em residual\/} of $f$.
\item[Fact 2:] $f$ and $g$ are monotone mappings, and there holds $g\circ f \geq {\rm id}_A$ and $f\circ g \leq {\rm id}_B$. Conversely, if two monotone mappings $f$ and $g$ satisfy $g\circ f \geq {\rm id}_A$ and $f\circ g \leq {\rm id}_B$, then they will form a residuated pair.
\item[Fact 3:] $f\circ g \circ f  = f$ and $g\circ f \circ g = g$. For this reason, the mappings $h:=g\circ f$ and $k:=f\circ g$ form {\em closure\/} and {\em kernel\/} operators, respectively. We observe that $h(x) = x$ for all $x\in {\rm im}(g)$ and $k(y) = y$ for all $y\in  {\rm im}(f)$. 
\item[Fact 4:] The mapping $f$ restricts to a bijection between the sets of closed elements in $A$ and kernel elements in $B$; this restriction is inverted by the restriction of  $g$ to the set of kernel elements in $B$. 
\item[Fact 5:] If $A$ and $B$ are complete lattices, then $f$ is residuated if and only if $f$ is preserving suprema, i.e.~$f(\sum X) = \sum f(X)$ for all $X \subseteq A$. Accordingly $g$ is a residual mapping if and only if it preserves infima, meaning $g(\prod Y) = \prod g(Y)$ for all $Y \subseteq B$.
\item[Fact 6:] Any residuated mapping $f: {\mathbb B}_2^n \longrightarrow {\mathbb B}_2^k$ can be canonically represented by an $n\times k$ matrix $H_f$ with entries in ${\mathbb B}_2$, where $f(x) = xH_f$ for all $x\in {\mathbb B}_2^n$. The representation of the residual mapping $g: {\mathbb B}_2^k \longrightarrow {\mathbb B}_2^n$ is the subject of the following theorem.
\end{description}

\smallskip

\begin{theorem}\label{represidual}
Let $f:{\mathbb B}_2^n \longrightarrow {\mathbb B}_2^k$ be a residuated mapping represented by the $n\times k$-matrix $H$.  Then the residual mapping $g: {\mathbb B}_2^k \longrightarrow {\mathbb B}_2^n$ is given by the assignment $y \mapsto \overline{\overline{y}H^T}$.
\end{theorem}
\begin{IEEEproof} In the following we will need to carefully observe the effects of the negation-operator: they will turn sums into products and products into sums. According to our list of facts, we only have to check if $g$ is monotone, and if $f\circ g \leq {\rm id}_{{\mathbb B}_2^k}$ and $g\circ f \geq {\rm id}_{{\mathbb B}_2^n}$. 

To begin, we compute \begin{eqnarray*}
[g (y)]_i & = & \overline{\sum_{j\leq k}\overline{y_j}H_{ij}} \; = \; \prod_{j\leq k}(\overline{H_{ij}} + y_j),
\end{eqnarray*} 
which already implies the monotone property of $g$.

For $y=f(x)$ we now have $y_j = \sum\limits_{\ell\leq n} x_\ell H_{\ell j}$, and this leads to \begin{eqnarray*}
[g\circ f(x)]_i & = & \prod_{j\leq k}(\overline{H_{ij}} + \sum_{\ell\leq n} x_\ell H_{\ell j}) \\ & \geq & \prod_{j\leq k}(\overline{H_{ij}} + x_i H_{ij})\\ & \geq &  \prod_{j\leq k}(x_i \overline{H_{ij}} + x_i H_{ij})
\\ & = & \prod_{j\leq k}x_i(\overline{H_{ij}}+ H_{ij}) \; =  \; x_i,
\end{eqnarray*}
for all $1 \leq i \leq n$, and thus $g\circ f(x) \geq x$ for all $x\in {\mathbb B}_2^n$.
Likewise, in a strictly dual fashion, we compute \begin{eqnarray*}
[f\circ g(y)]_j & = & \sum_{i\leq n} [g(y)]_i H_{i j}\\ & = & \sum_{i\leq n}  (\prod_{\ell\leq k}(\overline{H_{i \ell}} + y_\ell))H_{i j} \\
& \leq & \sum_{i\leq n}  (\overline{H_{ij}} + y_j) H_{ij} \\ & = & \sum_{i\leq n} y_j H_{ij} \; \leq \; y_j,\end{eqnarray*}
for all $1\leq j \leq k$, and hence $f\circ g(y) \leq y$ for all $y\in {\mathbb B}_2^k$, which completes the proof.
\end{IEEEproof}

\smallskip

We will shed further light on condition {\bf d--Rev}. A proof of the following theorem will be given in a more detailed account. 

\smallskip

\begin{theorem}
Let $f: {\mathbb B}_2^n \longrightarrow {\mathbb B}_2^k$ be a residuated mapping represented by the $n\times k$-matrix $H$, and let $g$ denote the residual mapping of $f$. Then the following are equivalent:
\begin{abc}
\item $H$ satisfies {\bf d--Rev}.
\item $g\circ f(x) = x$ for all $x \in B_\delta(\mathbf{0},d)$.
\item $B_\delta(\mathbf{0},d) \subseteq {\rm im}(g)$.
\item $B_\delta(\mathbf{1},d) \subseteq {\rm colspace}(H)$.
\end{abc}
\end{theorem}

\section{Group Testing}

We enter the central topic of the paper. We will follow the matrix model of non-adaptive group testing as first presented in~\cite{katona73} and reformulate it to suit our language of Residuation. 

\smallskip

\begin{definition}
Let $n$ and $k$ be natural numbers. An $(n,k)$-{\em group testing scheme\/} is a residuated mapping $f: {\mathbb B}_2^n \longrightarrow {\mathbb B}_2^k$. The matrix $H$ representing $f$ is referred to as {\em testing matrix\/}.
\end{definition}

\smallskip

As described in the introduction, typically an (unknown) vector $x\in {\mathbb B}_2^n$ represents a list of samples. These samples are distributed to $k$ different tests, and $H$ describes which sample is participating in which test. The testing process is mathematically represented by the multiplication of $x$ by $H$, so that $y = xH \in {\mathbb B}_2^k$ becomes the vector of (known) test results.   

It would now be desirable to have a mapping $g: {\mathbb B}_2^k\longrightarrow {\mathbb B}_2^n$ that assigns each test result $y\in {\mathbb B}_2^k$ a sample vector $x' \in {\mathbb B}_2^n$ (hopefully) showing a valid pattern of infective samples. Such a mapping $g$ would be referred to as a {\em decision scheme\/}, or a {\em decoder\/} for the group testing scheme.

\smallskip

\begin{definition}
An $(n,k,d)$-group testing scheme is an $(n,k)$-group testing scheme $f$ together with a decoder $g: {\mathbb B}_2^k \longrightarrow {\mathbb B}_2^n$ such that $g\circ f (x) = x$ for all $x\in B_\delta(\mathbf{0},d)$. 
\end{definition}

\smallskip

If such a decoder $g$ exists, we will say, $f$ allows the identification of up to $d$ infected samples, and it is obvious, that maximizing $d$ and minimizing $k$ are conflicting goals. For given $d\leq n$, an $(n,k,d)$ group testing scheme will be called {\em optimal\/}, if for every $(n,k',d)$ group testing scheme there holds $k'\geq k$.

There is a natural interest in decision schemes, because for group testing these are what syndrome decoders are for linear codes. 

\smallskip

\begin{theorem}
Let $f$ be an $(n,k,d)$-group testing scheme with testing matrix satisfying {\bf d--Rev}. Then the residual mapping $g$ of $f$ is a decoder for $f$. 
\end{theorem}

\section{Partial linear spaces}

In what follows, an {\em incidence structure\/} will be a pair $(P,B)$, where $P$ is a set of points, and where $B\subseteq 2^P$ is called the set of blocks. If a point $p\in P$ is contained in the block $c\in B$, then we say that $p$ is {\em incident\/} with $c$. 

All of the incidence structures in this paper will be finite, meaning $P$ (and hence $B$) are finite sets.

\smallskip 

\begin{definition}\label{incidencematrix}
Let $(P,B)$ be an incidence structure on the $v$-element set $P$ of points, and let $b=\left| B \right|$ denote the number of blocks of $B$. A binary matrix $M\in {\mathbb B}_2^{v \times b}$ is called an {\em incidence matrix\/} for $(P,B)$, if its rows are labelled by the points in $P$, while its columns are labelled by the blocks in $B$, such that $$M_{p,c} \; = \;  \left\{\begin{array}{ccl}
1 & : & p \in c,\\
0 & : & \mbox{otherwise.}
\end{array}\right.$$  
\end{definition}

\smallskip

Incidence matrices may thus be considered as indicator functions of their underlying incidence relation.

\smallskip

\begin{definition}
For natural number $s$ and $t$, a finite incidence structure $(P,L)$ consisting of {\em points\/} and {\em lines\/} is called a {\em partial linear space\/} of order $(s,t)$ if the following
axioms hold:
\begin{itemize}
	\item Every line  is incident with $s+1$ points, and every point
	is incident with $t+1$ lines.
	\item Two different points are connected by at most one line.
\end{itemize}
\end{definition}

\smallskip

It is obvious, that in a partial linear space, two different lines meet in at most one point, and hence this class of incidence structures is self-dual in the sense, that interchanging the terms ``line'' and ``point'' will transform a partial linear space of order $(s,t)$ into a partial linear space of order $(t,s)$.

A particularly well understood class of partial linear spaces is that of the {\em generalized quadrangles\/}. These spaces were first introduced by J.~Tits~\cite{Tit1} and enjoy cross links with many other structures in geometry and algebra in general.

\smallskip

\begin{definition}
A partial linear space $(P,L)$ of order $(s,t)$ is called a {\em generalized quadrangle\/}, denoted by $GQ(s,t)$, if for any non-incident point-line pair $(p,\ell)$, there exists a unique point $q$ on $\ell$ that is connected with $p$ by a line.
\end{definition}

\begin{figure}[ht]
	\begin{center}
		\includegraphics[height=4cm]{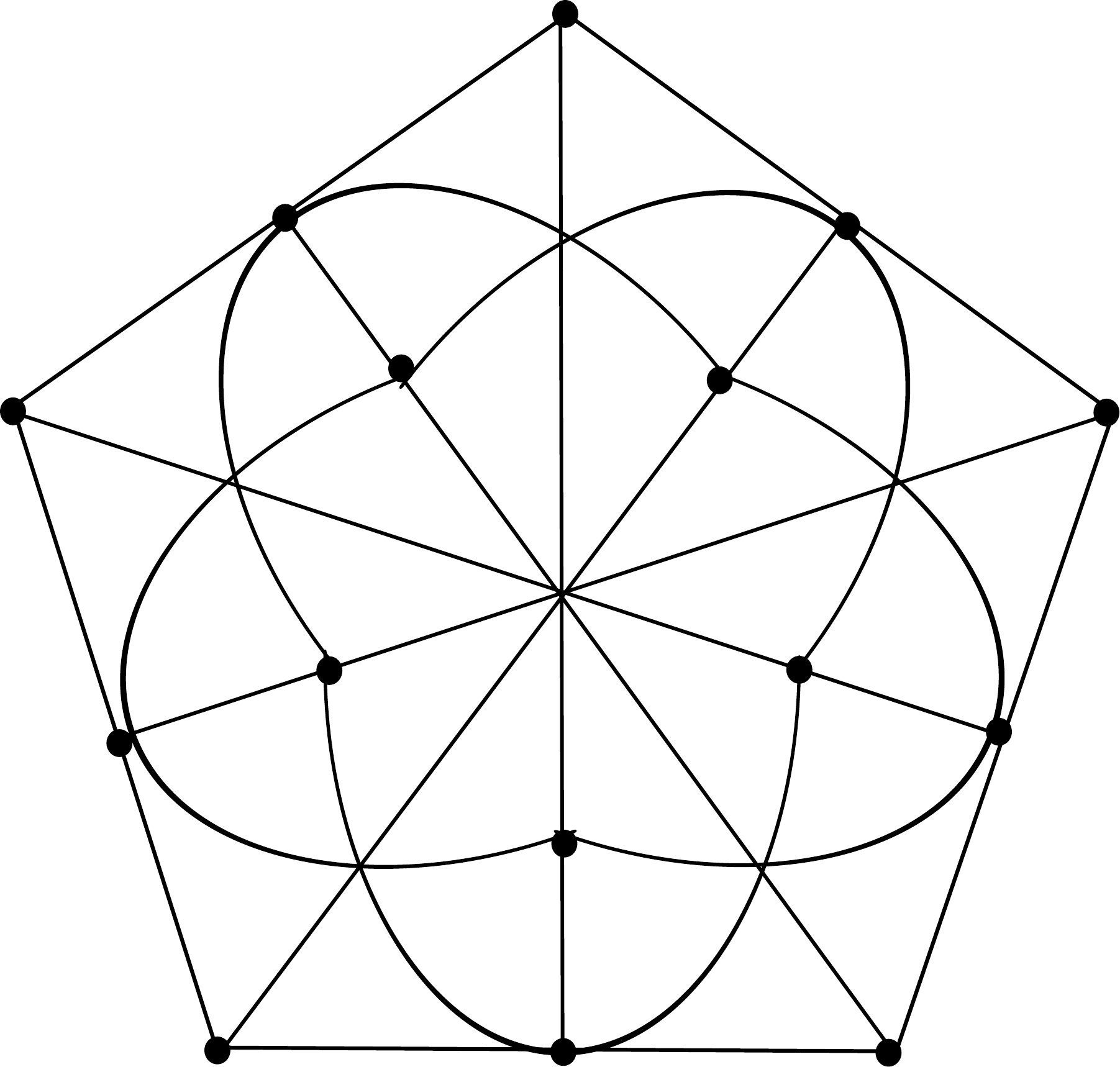}
		\caption{GQ(2,2) also known as W(2)}
		\label{FigGQ}
	\end{center}
\end{figure}

As indicated earlier, the pair $(s,t)$ is referred to as the {\em order} of the quadrangle. 

\smallskip

\begin{remark}
	A generalized quadrangle of order $(s,t)$ has $(s+1)(st+1)$ points and $(t+1)(st+1)$ lines.
\end{remark}

\smallskip

We refer to the standard reference \cite{PT} for more information on generalised quadrangles.

In terms of group testing, we can state the following theorem, the proof of which is rather simple.

\smallskip

\begin{theorem}
Let $(P,L)$ be a partial linear space of order $(s,t)$, and let ${\ell}_{1},\ldots {\ell}_{m}$ denote a collection of $m$ distinct lines in $L$. If ${\ell} \in L$ is a line with ${\ell} \subseteq  {\ell}_{1} \cup \cdots \cup {\ell}_m$ then ${\ell} = {\ell}_{j}$ for some $1 \leq j \leq m$ provided $m\leq s$.
\end{theorem}
\begin{IEEEproof} Assume $\ell \neq \ell_i$ for all $i=1, \ldots, m$. Then $|\ell \cap {\ell}_{i}| \leq 1$ which shows $$s+1\; = \; |\ell| \; \leq \; \Big| \ell \cap \bigcup_{i=1}^m \ell_i\Big|  \; \leq \; \sum_{i=1}^m \left|\ell \cap \ell_i\right| \; \leq \; m,$$ which contradicts $m \leq s$.
\end{IEEEproof}

\smallskip

For the incidence matrix of a partial linear space $(P,L)$ we may derive the following immediate conclusion.

\smallskip

\begin{corollary}
The incidence matrix of a partial linear space of order $(s,t)$ satisfies condition {\bf s--Rev}.
\end{corollary}

\smallskip

\begin{corollary}
The incidence matrix of a generalized quadrangle of order $(s,t)$ yields an $(n,k,d)$-group testing scheme where $n=(s+1)(st+1)$, $k=(t+1)(st+1)$, and $d=s$.
\end{corollary}

\section{Conclusion and Outlook}

We have presented the well-established discipline of non-adaptive group testing in the previously neglected language of Residuation Theory on vector spaces over the Boolean semifield ${\mathbb B}_2$. 

In doing so, we have developed a few novel aspects:

\smallskip

\begin{abc}
\item Our group testing schemes are modelled by residuated mappings $f$, the residual $g$ may serve as decision schemes (decoders) for the syndrome decoding problem: For given $y$ find $x$ such that $f(x)=y$.
\item These testing schemes identify every infection pattern that is contained in ${\rm im}(g)$. 
\item A condition {\bf d--Rev},  is introduced that exactly describes the condition under which $B_\delta(\mathbf{0},d) \subseteq {\rm im}(g)$. Exactly the group testing schemes satisfying {\bf d--Rev} will identify up to $d$ infected samples out of a given group of samples by application of the residual mapping $g$ as decoder.
\item If the matrix $H$ represents $f$, then {\bf d--Rev} is given for $H$ if and only if $B_\delta(\mathbf{1},d) \subseteq {\rm colspace}(H)$. 
\item Condition {\bf d--Dis} as discussed in \cite{essential} is equivalent to {\bf d--Rev} in this paper.
\item The incidence matrices of partial linear spaces of order $(s,t)$ satisfy {\bf s--Dis} and hence are suitable as testing matrices.
\item We have not made any assumptions on the relationship between $k$ and $n$. In traditional noise-free group testing, $k$ will be desired to be significantly smaller than $n$. In a noisy environment, and if error-correction is required as tests are inexpensive but unreliable, it appears possible to still strive for $k\leq n$, but in general we might need to allow for $k\geq n$.
\end{abc}

\section*{Acknowledgment}

We thank Ruth Greferath (Ph.D.) and Paul Wright (Ph.D.) for useful advice and discussions.

\end{document}